\documentclass[%
 aip,
 jmp,%
 amsmath,amssymb,
 reprint,%
]{revtex4-1}

\usepackage{natbib}
\usepackage{mathtools}
\usepackage{graphicx}% Include figure files
\usepackage{epstopdf}

\usepackage{dcolumn}% Align table columns on decimal point
\usepackage{bm}% bold math
\newcommand{\bv}[1]
{\mbox{\boldmath${#1}$}}

\begin{document}

\preprint{AIP/123-QED}

\title[
]{Dynamics of single-domain magnetic particles at elevated temperatures}

\author{M. Tzoufras}
 \email{mtzoufras@physics.ucla.edu}
\author{G. J. Parker}%
\author{M. K. Grobis}%
 \affiliation{HGST, a Western Digital Company, San Jose Research Center, USA}

\date{\today}

\begin{abstract} 
A stochastic differential equation that describes the dynamics of single-domain magnetic particles at any temperature is derived using a classical formalism. The deterministic terms recover existing theory and the stochastic process takes the form of a mean-reverting random walk. In the ferromagnetic state diffusion is predominantly angular and the relevant diffusion coefficient increases linearly with temperature before saturating at the Curie point ($T_c$). Diffusion in the macrospin magnitude, while vanishingly small at room temperature, increases sharply as the system approaches $T_c$. Beyond $T_c$, in the paramagnetic state, diffusion becomes isotropic and independent of temperature. The stochastic macrospin model agrees well with atomistic simulations.
\end{abstract}

%Use showkeys class option if keyword
\maketitle

The dynamics of single-domain magnetic particles at elevated temperatures are a fundamental physics problem that encompasses a phase transition at the Curie point ($T_c$). While the behavior near $T_c$ can be described using atomistic\cite{Ising:1925kx,Watson:1969uq,Nowak:2005ys,Evans:2014vn}  and renormalization\cite{RVictora} methods, more efficient approaches are needed for modeling macroscopic systems and for developing technologies based on magnetic phase transitions. The Landau-Lifshitz-Bloch (LLB) equation, derived by ensemble averaging a thermal distribution of interacting atomic spins using  either a quantum-mechanical\cite{Garanin:1991fk} or a classical\cite{Garanin:1997uq} framework, bridges the gap between room temperature, where a ferromagnet is described by the Landau-Lifshitz equation\cite{LandauLifshitz}, and temperatures above $T_c$, where the material becomes paramagnetic and its behavior is modeled by the Bloch equation\cite{Bloch:1946fk}. Nevertheless, the deterministic LLB equation is insufficient to capture the dynamics of single domain magnetic particles, because a degree of stochasticity---owing to the thermal fluctuations of individual atomic spins---survives  and plays a critical role in determining temporal evolution.

Thermal fluctuations in single-domain magnetic particles were first studied by augmenting the external magnetic field in the Landau-Lifshitz-Gilbert equation with a stochastic term\cite{Brown:1963fk}, thereby introducing randomness in the angle of the macrospin but not in its magnitude. The validity of this approach is limited to temperatures much lower than $T_c$, where the individual atomic spins are locked together by exchange interactions and the magnitude of the resulting macrospin is constant. Recently, stochastic fields were incorporated into the LLB equation in a number of different ways\cite{Garanin:2004uq, Evans:2012vn, Xu:2012fk, XuZhang2, GJZhu} in order to include the effect of thermal fluctuations at elevated temperatures. These formulations are distinct, the physics they describe are different,  and they exhibit markedly diverse stochastic behaviors when implemented in simulation codes.  

In this letter we present an improved stochastic LLB equation derived by considering the required behavior of macrospin thermal fluctuations. First we find the macrospin probability distribution for a system of  atomic spins in thermal equilibrium by invoking the mean field approximation and the central limit theorem.  We then formulate an advection-diffusion equation such that this probability distribution is the stationary solution and deduce the corresponding stochastic differential equation. We benchmark the new stochastic  equation against 
%Tzoufras
%mean field
%Tzoufras
 atomistic LLG calculations and find excellent agreement.

 In a classical framework\cite{Garanin:1997uq} an atomic spin is described by a unit vector $\bv{S}$ with magnetic moment $\bv{\mu} = \mu_0\bv{S}$. Under the influence of thermal fluctuations the spin direction at any point in time can vary, and we may use a distribution function $f(\bv{s})$  on the unit sphere $\vert \bv{s}\vert = 1$ to represent it. When an isolated classical spin is in thermal equilibrium, the function $f(\bv{s})$ takes the form of a Boltzmann distribution  $f_0(\bv{s})\propto \exp[-\mathcal{H}(\bv{s})/(k_BT)]$, where $\mathcal{H}(\bv{S}) = -\mu_0 \bv{H}\cdot\bv{S}$ is the Hamiltonian and $\bv{H}$ the total magnetic field. Using the normalized magnetic field $\bv{\xi}_0=\mu_0 \bv{H}/(k_B T)$ and assuming, without loss of generality, that the magnetic field is along the $z$-axis, i.e.  $\hat{\bv{H}} \equiv \hat{\bv{e}}_z$, we rewrite the equilibrium distribution as:
\begin{equation}
f_0(\bv{s})= \frac{\exp(\xi_0s_z)}{4\pi\sinh(\xi_0)/\xi_0}
\end{equation}

To calculate the moments of $f_0$ on the unit sphere we write  $\bv{s} $  in  spherical coordinates $ (s_x,s_y,s_z) = (\sin\vartheta\cos\varphi, \sin\vartheta\sin\varphi,\cos\vartheta)$ with $0\leq \varphi<2\pi$ and $0\leq \vartheta\leq \pi$. The averages $\langle S_x\rangle$ and $\langle S_y\rangle$, as well as the covariances $\langle S_iS_{j}\rangle_{i\neq j}$, vanish identically, and we obtain the average $\langle \bv{S}\rangle$ along the $z$-axis and a diagonal covariance matrix $\bv{\Sigma}=diag(\sigma_\perp^2,\sigma_\perp^2,\sigma_\parallel^2)$:
\begin{eqnarray}\label{averagem0}
\bv{m}_0&\equiv& \langle \bv{S}\rangle=\langle S_z\rangle \hat{\bv{e}}_z= \Bigl[\coth(\xi_0) -\frac{1}{\xi_0}\Bigr]\hat{\bv{H}}
\\
 \label{sigperp}
\sigma_\perp^2 &\equiv& Var(S_x)= Var(S_y) =  \frac{m_0}{\xi_0} 
\\
 \label{sigpar}
\sigma_\parallel^2 &\equiv& Var(S_z) = 1- m_0^2  - \frac{2m_0}{\xi_0} = \frac{\partial m_0}{\partial \xi_0}
\end{eqnarray}

The normalized magnetization of a single-domain particle comprised of $n$ unit cells is $\bv{M}^{(n)}=\frac{1}{n} \sum_{i=1}^n \bv{S}^{(i)}$.  To render  the random vectors $\bv{S}^{(i)}$ independent from one another we drop the spin-spin exchange interactions in favor of a mean field (mean field approximation). Moreover, we assume that both the  mean field and temperature are constant throughout the magnetic particle such that $\langle\bv{S}^{(i)}\rangle = \bv{m}_0$ and $\bv{\Sigma}^{(i)} = \bv{\Sigma}$ for all unit cells.  If the number of unit cells is large enough their average follows the multivariate normal distribution with $\sqrt{n}(\bv{M}^{(n)}-\bv{m}_0 )\xrightarrow{D}\mathcal{N}_{\{x,y,z\}}(0,\bv{\Sigma})$, according to the central limit theorem,  regardless of the individual $\bv{S}^{(i)}$-distributions. We may thus write the macrospin probability distribution:
\begin{equation}\label{pdfeq}
F_0(\bv{m}) = \frac{\exp\biggl\{-\frac{1}{2}\Bigl[\frac{(m_z-m_0)^2}{\sigma_\parallel^2/n}+\frac{m_x^2+m_y^2}{\sigma_\perp^2/n}\Bigr]\biggr\}}{(2\pi/n)^{3/2}\sigma_\parallel\sigma_\perp^2}
\end{equation}

We seek the stochastic process for which $F_0$ from Eq. (\ref{pdfeq}) is the stationary solution. To develop the relevant advection-diffusion equation, i.e. $\partial_{\tau} F = -\partial_{\bv{m}} \cdot[ \bv{\upsilon} F -  \mathfrak{D}\cdot \partial_{\bv{m}} F] $, we need the advective ($\bv{v}F$) and diffusive ($\mathfrak{D}\cdot \partial_{\bv{m}} F$) fluxes to cancel out when the system is stationary ($F\equiv F_0$). By calculating $\partial_{\bv{m}}F_0$ we immediately find  $(\bv{m}-\bv{m}_0)F_0 + \frac{1}{n} \bv{\Sigma}\cdot\partial_{\bv{m}} F_0=0$ and identify the first term as the advective flux and the second as the diffusive flux. The conservation law can be written as:
\begin{equation}\label{adeq}
\frac{\partial F}{\partial t}\biggr\vert_{\mbox{adv-diff}}  = \frac{1}{\tau_s}\frac{\partial}{\partial \bv{m}}\cdot \Bigl[ (\bv{m}-\bv{m}_0)F + \frac{\bv{\Sigma}}{n} \cdot\frac{\partial F}{\partial  \bv{m}} \Bigr]
\end{equation}
 with $\tau_s$ a coefficient in units of time.

For a system rotating around a magnetic field $\bv{H}$ we can write the rotation flux $-(\bv{m}\times \bv{H}) F$, which is divergenceless when $F=F_0$.  The resulting precession equation is: 
 \begin{equation}\label{roteq}
 \frac{\partial F}{\partial t}\biggr\vert_{\mbox{rotation}}  = \gamma\frac{\partial}{\partial \bv{m}}\cdot [ (\bv{m}\times \bv{H}) F]
 \end{equation} 
and we recognize $\gamma$ as the gyromagnetic ratio with $\gamma\mu_0\bv{S}$ the mechanical moment of an atomic spin.

 The final Fokker-Planck equation for the distribution $F$ combines both the stochastic [Eq. (\ref{adeq})] and precession [Eq. (\ref{roteq})] processes:
 \begin{equation}\label{fpeq1}
\frac{\partial F}{\partial t} =\frac{\partial}{\partial \bv{m}}\cdot\biggl[\gamma(\bv{m}\times \bv{H})F+\frac{1}{\tau_s}\Bigl(\bv{m}-\bv{m}_0 + \frac{\bv{\Sigma}}{n}\cdot\frac{\partial}{\partial\bv{m}}\Bigr)F \biggr]
\end{equation}

The second term on the right-hand-side (RHS) of Eq. (\ref{fpeq1}) corresponds to an Ornstein-Uhlenbeck (OU) stochastic process with stationary solution $F_0$ in Eq. (\ref{pdfeq}). This process describes a mean-reverting random walk, a typical example of which is the velocity of a massive Brownian particle under the influence of friction. Here the stochasticity is due to the finite number ($n$) of unit cells contained in a magnetic grain, and $\tau_s^{-1}$ incorporates an effective friction coefficient\cite{Neel}. For an infinitely large particle ($n\rightarrow \infty$) the diffusive flux vanishes and $F_0$ becomes a delta function. The diffusional relaxation time $\tau_s$ is related to the phenomenological damping parameter ($\lambda$) in the LLG equation (see below) and is a function of both temperature and material properties.

In order to simplify Eq. (\ref{fpeq1}) we express the tensor product in terms of vectors perpendicular and parallel to the axis imposed by the total magnetic field $\hat{\bv{H}}\equiv \hat{\bv{e}}_z$ (which includes exchange, magnetic anisotropy and externally applied fields)  to obtain:
\begin{widetext}
\begin{equation}\label{longFP}
\frac{\partial F}{\partial t} =-\frac{\partial}{\partial \bv{m}}\cdot\biggl\{-\gamma(\bv{m}\times \bv{H})F+\frac{1}{\tau_s}\Bigl[ -(\bv{m}-\bv{m}_0)+ \frac{\sigma_\perp^2\mu_0}{M_s^0 V} \hat{\bv{H}}\times\Bigl(\hat{\bv{H}}\times \frac{\partial F}{\partial \bv{m}}\Bigr)- \frac{\sigma_\parallel^2\mu_0}{M_s^0 V} \Bigl(\hat{\bv{H}}\cdot \frac{\partial F}{\partial \bv{m}}\Bigr)\hat{\bv{H}}  \Bigr] \biggr\}
\end{equation}
\end{widetext}
where $n$ has been substituted by $M_s^0V/\mu_0$, with $M_s^0$ the saturation magnetization per unit volume at $T=0K$. The first two terms on the RHS of Eq. (\ref{longFP}) yield the deterministic dynamics and the last two terms the diffusive behavior.  The third and forth terms describe the stochasticity in terms of transverse and longitudinal diffusion with respect to the  $\hat{\bv{H}}$-axis, with diffusion coefficients $D_{\perp, \parallel }^{H}= \frac{\sigma_{\perp, \parallel}^2\mu_0}{M_s^0 V}$, rather than the $\hat{\bv{m}}$-axis as in Refs. \cite{Garanin:2004uq, Evans:2012vn, Xu:2012fk, XuZhang2}.  Below $T_c$, where the exchange field dominates, the macrospin  points roughly along the equilibrium direction $\hat{\bv{m}} \simeq \hat{\bv{m}}_0 \equiv \hat{\bv{H}}$, and the distinction between the decomposition used here and the one in the literature disappears. Above $T_c$, where $D_\parallel^{H}(T>T_c)\simeq D_\perp^{H}(T>T_c)$, the decomposition along and across $\hat{\bv{H}}$ is redundant. Consequently, the differentiation between $\hat{\bv{m}}$ and $\bv{\hat{H}}$ is only meaningful in the immediate vicinity of the Curie temperature.

A numerical scheme that solves Eq. (\ref{longFP}) would have to keep track of the three-dimensional probability distribution $F(\bv{m})$. To formulate a simpler and less computationally intensive approach we write a stochastic differential equation in accordance with the Fokker-Planck equation [Eq. (\ref{longFP})]:
\begin{widetext}
\begin{equation}\label{sODE}
\frac{d\bv{m}}{dt} =  -\gamma (\bv{m}\times \bv{H})-\frac{1}{\tau_s}(\bv{m} -\bv{m}_0) +  \frac{\sigma_\parallel}{\sigma_\perp} ( \hat{\bv{H}} \cdot \bv{\eta})\hat{\bv{H}} - \hat{\bv{H}} \times (\hat{\bv{H}} \times \bv{\eta}),\qquad
    \langle\eta_i(t_1)\eta_j(t_2)\rangle = 2\frac{\sigma_{\perp}^2}{\tau_s n} \delta_{ij} \delta(t_2-t_1)
\end{equation}
\end{widetext}
 where $\bv{m}_0 = m_0\hat{\bv{H}}$, and the coefficients $\sigma_\parallel,\sigma_\perp$ are given in  Eqs. (\ref{averagem0})-(\ref{sigpar}). At room temperature, where $ \hat{\bv{m}} \simeq \hat{\bv{H}}$ and $\sigma_\parallel\ll \sigma_\perp$, Eq. (\ref{sODE}) reduces to an LLB equation with purely angular diffusion and it recovers the stochastic behavior in Ref. \cite{Brown:1963fk}. For $T>T_c$ we obtain $\sigma_\parallel^2 \simeq \sigma_\perp^2 \simeq 1/3$ and the last two terms on the RHS of Eq. (\ref{sODE}) can be combined to yield the Langevin field $\bv{\eta}$. 

To further elucidate the diffusive behavior we substitute the normalized magnetic field $\xi_0$ with $\mu_0 H/ (k_B T)$ and rewrite the diffusion coefficient $D_{\perp}^{H}= \frac{m_0}{H} \frac{k_B T}{M_s^0 V}$. This is identical to the angular diffusion coefficient in Ref. \cite{Xu:2012fk}, such that angular diffusion in Eqs. (\ref{longFP})-(\ref{sODE}) and in Ref. \cite{Xu:2012fk} are the same everywhere except near $T_c$, where $\hat{\bv{m}}(T_c) \not\approx \hat{\bv{H}}(T_c)$. To obtain an expression for the temperature dependence of the diffusion coefficient $D_{\perp}^{H}$ we write the total magnetic field as $\bv{H}\simeq (3k_BT_c/\mu_0) \bv{m} +\bv{H}_{\mbox{eff}} $, where $\bv{H}_{\mbox{eff}}$ includes the anisotropy and external fields. Below $T_c$ the exchange field dominates, $\mu_0\vert \bv{H}_{\mbox{eff}}\vert\ll3k_BT_c\vert \bv{m}\vert$,  and the angular diffusion is a linear function of temperature $D_{\perp}^{H}(T<T_c)\propto T/(3T_c)$. For temperatures above $T_c$, $\coth(\xi_0\ll 1) = \xi_0^{-1}+\xi_0/3-\ldots \Rightarrow m_0/\xi_0 \simeq 1/3$ and we can write $\sigma_\perp^2(T>T_c) \simeq 1/3$. Physically, above $T_c$ the exchange field vanishes and (assuming $\bv{H}_{\mbox{eff}} \simeq 0$) atomic spins have no preferred direction, such that the atomic spin distribution  
%Tzoufras
%$f_0(\bv{s})$
%Tzoufras 
 fills the entire spherical shell $\vert\bv{s}\vert \equiv 1$ homogeneously. The resulting variance is then identical in all directions  $Var(S_{\{x,y,z\}}) (T> T_c) = \int_{4\pi}\cos^2\vartheta d\Omega = 1/3$. An approximate expression for $D_{\perp}^{H}$ with temperature may be written as:
\begin{equation}
D_{\perp}^{H} \simeq \frac{\mu_0}{3 M_s^0V} \times
\left\{\begin{array}{c}
T/T_c, \qquad T\leq T_c\\
1, \qquad T>T_c
\end{array}
\right.
\end{equation}

The diffusion in the macrospin magnitude $D_\parallel^{H}$ saturates at the same value as $D_\perp^{H}$ above $T_c$. However, below the Curie point, $D_\parallel^{H}$ is not a linear function of temperature. Instead, its properties can be found by using the expression $\sigma_\parallel^2 = \partial m_0/\partial \xi_0 \equiv L^\prime(\xi_0)$, where  $L(\xi_0) = \coth(\xi_0)-1/\xi_0$ is the Langevin function and the prime denotes differentiation with respect to $\xi_0$. For temperatures much lower than $T_c$ the slope of $m_0(T)$ is very small (especially if one considers the Brillouin \cite{kittelb} instead of the Langevin function), and as a result $D_\parallel^{H}(T\ll T_c)$ can be neglected when compared to $D_\perp^{H}(T\ll T_c)$.  As the temperature approaches $T_c$ from below, the diffusion in the magnetization magnitude $D_\parallel^{H}$ increases sharply following the magnitude of the slope of $m_0(T)$. At the Curie point, the slope of $m_0(T)$ vanishes abruptly and $D_\parallel^{H}$ saturates.  Beyond $T_c$  the system is isotropic and the overall diffusive flux in the Fokker-Planck equation [Eq. (\ref{longFP})] reduces to $(3M_s^0V/\mu_0)^{-1}\partial_{\bv{m}}F$. This is consistent with the physical picture of a paramagnetic grain, for which all directions are equivalent and the expected stochastic behavior is isotropic.

The deterministic terms of Eq. (\ref{sODE}) are identical to those in Ref.  \cite{Xu:2012fk} and may be expressed in the same form as Eq. (2.17) in Ref. \cite{Garanin:1997uq}:
\begin{equation}
\frac{d\bv{m}}{dt}= -\gamma (\bv{m}\times \bv{H})-\Gamma_1\Bigl(1 - \frac{\bv{m}_0\cdot\bv{m}}{m^2}\Bigr)\bv{m} -\Gamma_2\frac{\bv{m} \times (\bv{m}\times \bv{m}_0)}{m^2} 
\end{equation}
 
 Here we find $\Gamma_1 = \Gamma_2 = \tau_s^{-1}$, in contrast to the expression in Ref. \cite{Garanin:1997uq}, where  $\Gamma_1\equiv \gamma\frac{2\lambda k_BT}{\mu_0}\frac{m_0}{\xi_0 m_0^\prime}\neq\gamma\frac{2\lambda k_BT}{\mu_0}\frac{1}{2}(\frac{\xi_0}{m_0}-1)\equiv\Gamma_2$, with $\lambda$ the phenomenological damping parameter from the LLG-Langevin equation. This discrepancy arises because the present derivation, unlike the one in Ref. \cite{Garanin:1997uq}, does not assume that individual atoms obey LLG-Langevin dynamics. The classical treatment of dissipation in atomic spins, in which all dissipative processes (e.g. spin interactions with photons and spin waves) are approximated by a single damping term, has been employed in many theoretical and numerical investigations.  With that in mind, for $T\gtrsim T_c$, the expressions from Ref. \cite{Garanin:1997uq} yield $\Gamma_1(T\gtrsim T_c) \simeq \Gamma_2(T\gtrsim T_c) \simeq 2\gamma\lambda k_BT/\mu_0$ in agreement with the deterministic terms in Eq. (\ref{sODE}).  Hence, for magnetic particles near the Curie point,  we may set the relaxation rate to $\tau_s^{-1} \simeq \Gamma_1\simeq \Gamma_2$. 
 
To verify the macrospin model in Eq. (\ref{sODE}) we benchmarked it against atomistic simulations that solve the LLG-Langevin equation for every unit cell.  
We consider systems of FePt nanoparticles, which are of great technological interest especially for mangetic recording\cite{FePt1, Sun:2000qf},  with $fct$ crystal structure that comprises $2$ Fe atoms per unit cell of volume $v = 3.7\mbox{\AA}\times (3.88\mbox{\AA})^2$,  anisotropy energy density $K_1 = 7.64\times 10^7 erg/cm^3$, and  magnetic moment $\mu_0 = 3.23\mu_B$ ($\mu_B$ being the Bohr magneton). The total particle volume is set to $V = 7744v \simeq 431(nm)^3$, a mean field $\bv{H} = 3k_BT_c/\mu_0\langle \bv{S}\rangle + \bv{H}_{\mbox{eff}}$ is used, where $\bv{H}_{\mbox{eff}}$  includes the anisotropy ($\bv{H}_k$) and external ($\bv{H}_w$) fields, and we chose the phenomenological damping parameter for the LLG-Langevin equation $\lambda = 0.1$. 
%Tzoufras
The Curie temperature was set to $T_c=646K$ following fully atomistic simulations (below). 
%Tzoufras
 The same parameters are used for benchmarking the stochastic macrospin models in Refs. \cite{Evans:2012vn, Xu:2012fk} and Eq. (\ref{sODE}). We chose $\tau_s^{-1} = \Gamma_1$ both for Eq. (\ref{sODE}) and for the angular diffusion model from Ref. \cite{Xu:2012fk}. 

We examined the cooling of single-domain magnetic particles through $T_c$ for a wide range of parameters. For each parameter choice $4096$ particles are initialized at $750K$ and their temperature decays exponentially  $T[K] = 300 + 450\times \exp(-t/\tau)$. The benchmarks include two time constants, $\tau = 100ps$ and $\tau=1ns$, and a parameter scan over the external field magnitude ($H_w$) and its angle ($\theta$) with the anisotropy axis. In each simulation we measure the percentage of particles ($P^\uparrow$) for which $\bv{M}\cdot \bv{H}_w > 0$. At the end of the cooling process, when the macrospin is frozen, the final $P^\uparrow(T\ll T_c)$ represents the probability that a particle has been written correctly, i.e. the ``write probability".  Simulations use the Heun numerical scheme\cite{Garcia-Palacios:1998fk} with $0.5fs$ time-step.

\begin{figure}[htbp]
\includegraphics[width=8.8cm]{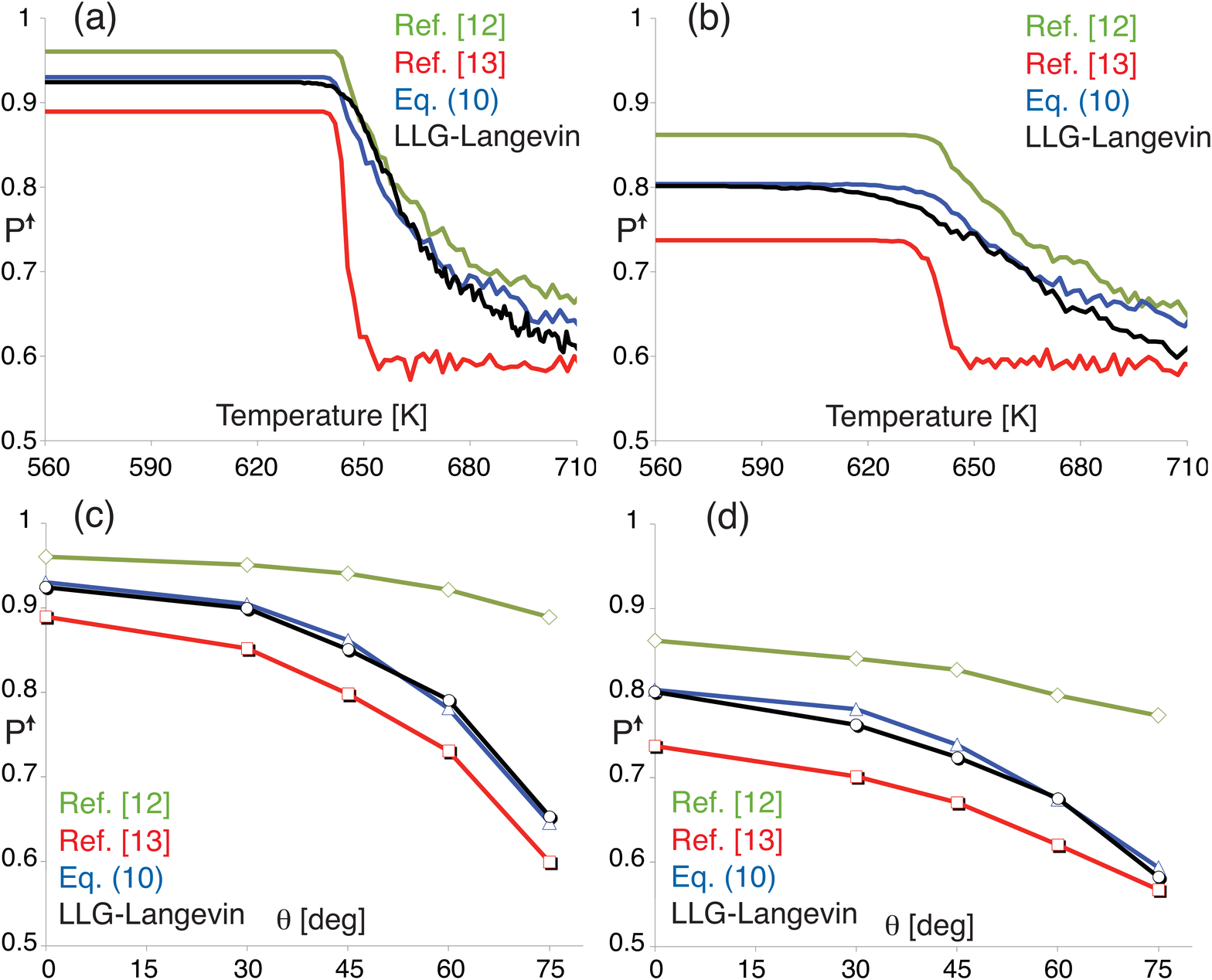}
\caption{\label{fig:comparemodels} 
Comparison between macrospin models and 
%Tzoufras
mean field 
%Tzoufras
 atomistic LLG-Langevin simulations. Results have been averaged over $4096$ particles. The percentage of particles ($P^\uparrow$) for which $\bv{M}\cdot \bv{H}_w > 0$ is shown as a function of temperature during a cooling process $T[K] = 300 + 450\times \exp(-t/\tau)$, with timescale $\tau = 1ns$ in panel (a), and $\tau = 100ps$ in panel (b). The write probability, i.e. $P^\uparrow(T\ll T_c)$, is shown as a function of field angle ($\theta$ measured at $0^\circ,30^\circ,45^\circ,60^\circ,75^\circ$) when the time-scale is $1ns$ in panel (c), and $100ps$ in panel (d). $H_w = 5kOe$ everywhere.} 
\end{figure}

 In Figure \ref{fig:comparemodels} panel (a) we compare the macrospin models, Ref.\cite{Evans:2012vn}/Ref.\cite{Xu:2012fk}/Eq. (\ref{sODE}) in  green/red/blue, with atomistic LLG-Langevin simulations (black) for $H_w = 5kOe, \theta = 0^\circ, \tau = 1ns$.  As the temperature decreases, the angular diffusion model from Ref.\cite{Xu:2012fk} (red) predicts virtually no change in $P^\uparrow$, until the Curie point is reached, where $P^\uparrow$ increases sharply before freezing. On the other hand, the stochastic LLB model from Ref. \cite{Evans:2012vn} (green) qualitatively recovers the smooth transition near $T_c$ seen from the atomistic simulations. The new stochastic LLB model in Eq. (\ref{sODE}) (blue) recovers the features of the atomistic simulations both qualitatively and quantitatively. The same behavior is observed in panel (b), where the time-scale $\tau$ is reduced to $100ps$. In this case the model from Ref. \cite{Evans:2012vn} (green) significantly deviates from the atomistic simulation. Panels (c) and (d)  show the dependence of the write probability, $P^\uparrow(T\ll T_c)$, with field angle  $\theta$, for the time-scales $\tau = 1ns$ and $\tau=100ps$ respectively. Good agreement between Eq. (\ref{sODE}) and the atomistic simulations is found for all cases.

\begin{figure}[htbp]
\includegraphics[width=8.8cm]{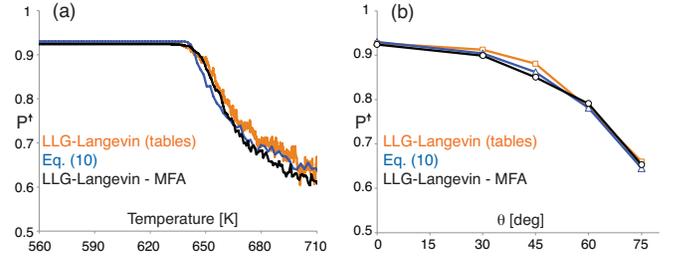}
\caption{\label{fig:compareatomistic} 
Comparison between the new macrospin model in Eq. (\ref{sODE}) and atomistic LLG-Langevin simulations, both with and without the mean field approximation. Results for the macrospin and mean-field atomistic simulations have been averaged over $4096$ particles. Results for the atomistic calculations without the MFA have been averaged over $1024$ particles. The percentage of particles ($P^\uparrow$) for which $\bv{M}\cdot \bv{H}_w > 0$ is shown as a function of temperature during a cooling process $T[K] = 300 + 450\times \exp(-t/\tau)$, with timescale $\tau = 1ns$ in panel (a). The write probability, i.e. $P^\uparrow(T\ll T_c)$, is shown as a function of field angle ($\theta$ measured at $0^\circ,30^\circ,45^\circ,60^\circ,75^\circ$) in panel (b). $H_w = 5kOe$ everywhere.} 
\end{figure}

The LLB equation is used extensively to model systems for magnetic data storage, such as Heat-Assisted Magnetic Recording (HAMR)\cite{HAMR2,Lyberatos:2003zr,HAMR1} and magneto-optical recording\cite{Stanciu:2007fk,Mangin:2014ly}. Current HAMR media employ FePt grains that contain more than $10^4$ unit cells. Even for future media with reduced grain-sizes, the number of unit cells can be expected to exceed $10^3$, sufficient to justify application of the central limit theorem in deriving the macrospin distribution [Eq. (\ref{pdfeq})].  However, the underlying assumption that the mean field and temperature are constant across the magnetic grain, is weaker. For example, for unit cells on the surface of the grain the overall magnetic field can be considerably smaller because there are fewer neighboring atoms to interact with. For systems where the number of surface unit cells is comparatively large, that is for small or irregularly-shaped grains, this effect becomes significant. Moreover, temperature variations within the grain influence the strength of the exchange and anisotropy fields, thereby undermining the accuracy of the macrospin model. The mean field approximation also precludes the excitation of internal modes of the grain that reflect the properties of the exchange interaction.  Atomistic models can be used to lift the restrictions due to the mean field approximation, and they offer the flexibility of modeling heterogenous magnetic materials, albeit at considerable computational cost. 
%Tzoufras 
%Benchmarking against such models is beyond the scope of the present work. 

To examine the applicability of the mean field theory we compare the new macrospin model [Eq. (\ref{sODE})] and the mean field atomistic simulations with a fully atomistic model  in Figure  \ref{fig:compareatomistic}. For the fully atomistic simulations the magnetic field $\bv{H}_i$  at each lattice point is calculated from the spin Hamiltonian: $\bv{H}_i = \partial \mathcal{H}/\partial \bv{S}_i $, where $\mathcal{H} = -\sum_{i\neq j} \bv{S}_i\cdot \bv{J}_{ij}^M\cdot\bv{S}_j -  \sum_{i}(k_{Fe}^{(0)}+m_i K^\prime)(\bv{S}_i\cdot \hat{\bv{e}}_z)^2 - \mu_0 \bv{H}_w\cdot \sum_i\bv{S}_i$. The tensor $\bv{J}_{ij}^M$ describes the exchange interactions with  $8\times8\times6\times2= 768$ terms, $k_{Fe}^{(0)} = -0.097meV$, $K^\prime = 0.0223meV$,  and $m_i$ is the number of nearest neighbors at each lattice point\cite{ALyberatos}. The simulation parameters and conditions were the same as those described in Figure \ref{fig:comparemodels} for the time-scale $\tau = 1ns$. The results from the fully atomistic simulations have been averaged over $1024$ particles and are shown in Figure \ref{fig:compareatomistic} with orange lines. Panels \ref{fig:compareatomistic}(a) and \ref{fig:compareatomistic}(b) correspond to \ref{fig:comparemodels}(a) and \ref{fig:comparemodels}(c)  and confirm that there is good agreement between Eq. (\ref{sODE}) and the fully atomistic simulations for all cases.
%Tzoufras

We have developed a stochastic Landau-Lifshitz-Bloch equation [Eq. (\ref{sODE})] for single-domain magnetic particles, in which the stationary state is determined directly from the underlying atomic spin equilibrium and the stochastic process is a mean-reverting random walk. At room temperature the stochastic behavior reduces to angular diffusion in agreement with the stochastic LLG equation. The angular diffusion coefficient increases linearly with temperature before saturating at $T_c$. Diffusion in the magnetization magnitude increases rapidly with temperature when $T\lesssim T_c$ and saturates at $T_c$. The stochastic LLB model is in excellent agreement with mean-field atomistic LLG-Langevin simulations and is immediately applicable for modeling a variety of magnetic  systems.
 
\bibliographystyle{unsrt}

\begin{thebibliography}{10}

\bibitem{Ising:1925kx}
Ernst Ising.
\newblock Beitrag zur theorie des ferromagnetismus.
\newblock {\em Zeitschrift f{\"u}r Physik}, 31(1):253--258, 1925.

\bibitem{Watson:1969uq}
R.~E. Watson, M.~Blume, and G.~H. Vineyard.
\newblock Spin motions in a classical ferromagnet.
\newblock {\em Physical Review}, 181(2):811--823, 05 1969.

\bibitem{Nowak:2005ys}
U.~Nowak, O.~N. Mryasov, R.~Wieser, K.~Guslienko, and R.~W. Chantrell.
\newblock Spin dynamics of magnetic nanoparticles: Beyond brown's theory.
\newblock {\em Physical Review B}, 72(17):172410--, 11 2005.

\bibitem{Evans:2014vn}
R~F~L Evans, W~J Fan, P~Chureemart, T~A Ostler, M~O~A Ellis, and R~W Chantrell.
\newblock Atomistic spin model simulations of magnetic nanomaterials.
\newblock {\em Journal of Physics: Condensed Matter}, 26(10), 2014.

\bibitem{RVictora}
R.~H. Victora and Pin-Wei Huang.
\newblock Simulation of heat-assisted magnetic recording using renormalized
  media cells.
\newblock {\em IEEE Trans. on Magn.}, 49(2):751--757, February 2013.

\bibitem{Garanin:1991fk}
D.~A Garanin.
\newblock Generalized equation of motion for a ferromagnet.
\newblock {\em Physica A: Statistical Mechanics and its Applications},
  172(3):470--491, 4 1991.

\bibitem{Garanin:1997uq}
D.~A. Garanin.
\newblock Fokker-planck and landau-lifshitz-bloch equations for classical
  ferromagnets.
\newblock {\em Physical Review B}, 55(5):3050--3057, 02 1997.

\bibitem{LandauLifshitz}
Lifshitz E.~M. Landau L.~D.
\newblock On the theory of the dispersion of magnetic permeability in
  ferromagnetic bodies.
\newblock {\em Phys. Zeitsch. der Sow.}, 8:153--169, 1935.

\bibitem{Bloch:1946fk}
F.~Bloch.
\newblock Nuclear induction.
\newblock {\em Physical Review}, 70(7-8):460--474, 10 1946.

\bibitem{Brown:1963fk}
William~Fuller Brown.
\newblock Thermal fluctuations of a single-domain particle.
\newblock {\em Physical Review}, 130(5):1677--1686, 06 1963.

\bibitem{Garanin:2004uq}
D.~A. Garanin and O.~Chubykalo-Fesenko.
\newblock Thermal fluctuations and longitudinal relaxation of single-domain
  magnetic particles at elevated temperatures.
\newblock {\em Physical Review B}, 70(21):212409, 12 2004.

\bibitem{Evans:2012vn}
R.~F.~L. Evans, D.~Hinzke, U.~Atxitia, U.~Nowak, R.~W. Chantrell, and
  O.~Chubykalo-Fesenko.
\newblock Stochastic form of the landau-lifshitz-bloch equation.
\newblock {\em Physical Review B}, 85(1):014433, 01 2012.

\bibitem{Xu:2012fk}
Lei Xu and Shufeng Zhang.
\newblock Magnetization dynamics at elevated temperatures.
\newblock {\em Physica E: Low-dimensional Systems and Nanostructures},
  45(0):72--76, 8 2012.

\bibitem{XuZhang2}
Lei Xu and Shufeng Zhang.
\newblock Self-consistent bloch equation and landau-lifshitz-bloch equation of
  ferromagnets: A comparison.
\newblock {\em Journal of Applied Physics}, 113(16):163911, 2013.

\bibitem{GJZhu}
J.-G. Zhu and Hai Li.
\newblock Understanding signal and noise in heat assisted magnetic recording.
\newblock {\em IEEE Trans. on Magn.}, 49(2):765--772, February 2013.

\bibitem{Neel}
L.~N{\'e}el.
\newblock Th{\'e}orie du tra{\^\i}nage magn{\'e}tique des ferromagn{\'e}tiques
  en grains fins avec applications aux terres cuites.
\newblock {\em Ann. G{\'e}ophys}, 5:99--136, 1949.

\bibitem{kittelb}
Kittel C.
\newblock {\em Introduction to Solid State Physics}.
\newblock Wiley, 8 edition, 2005.

\bibitem{FePt1}
D.~Weller, Andreas Moser, Liesl Folks, Margaret~E. Best, Wen Lee, M.F. Toney,
  M.~Schwickert, J.-U. Thiele, and Mary~F. Doerner.
\newblock High ku materials approach to 100 gbits/in2.
\newblock {\em 10 IEEE Transactions on Magnetics}, 36(1):10--15, January 2000.

\bibitem{Sun:2000qf}
Shouheng Sun, C.~B. Murray, Dieter Weller, Liesl Folks, and Andreas Moser.
\newblock Monodisperse fept nanoparticles and ferromagnetic fept nanocrystal
  superlattices.
\newblock {\em Science}, 287(5460):1989--1992, 03 2000.

\bibitem{Garcia-Palacios:1998fk}
Jos{\'e}Luis Garc{\'\i}a-Palacios and Francisco~J. L{\'a}zaro.
\newblock Langevin-dynamics study of the dynamical properties of small magnetic
  particles.
\newblock {\em Physical Review B}, 58(22):14937--14958, 12 1998.

\bibitem{HAMR2}
S.R. Cumpson, P.~Hidding, and R.~Coehoorn.
\newblock A hybrid recording method using thermally assisted writing and flux
  sensitive detection.
\newblock {\em IEEE Transactions on Magnetics}, 36(5):2271 -- 2275, September
  2000.

\bibitem{Lyberatos:2003zr}
A.~Lyberatos and K.~Yu Guslienko.
\newblock Thermal stability of the magnetization following thermomagnetic
  writing in perpendicular media.
\newblock {\em Journal of Applied Physics}, 94(2):1119--1129, 2003.

\bibitem{HAMR1}
R.E. Rottmayer, Sharat Batra, D.~Buechel, W.A. Challener, Julius Hohlfeld,
  Y.~Kubota, Lei Li, Bin Lu, C.~Mihalcea, Keith Mountfield, Kalman Pelhos,
  Chubing Peng, T.~Rausch, Michael~A. Seigler, D.~Weller, and XiaoMin Yang.
\newblock Heat-assisted magnetic recording.
\newblock {\em IEEE Transactions on Magnetics}, 42(10):2417--2421, October
  2006.

\bibitem{Stanciu:2007fk}
C.~D. Stanciu, F.~Hansteen, A.~V. Kimel, A.~Kirilyuk, A.~Tsukamoto, A.~Itoh,
  and Th. Rasing.
\newblock All-optical magnetic recording with circularly polarized light.
\newblock {\em Physical Review Letters}, 99(4):047601--, 07 2007.

\bibitem{Mangin:2014ly}
S.~Mangin, M.~Gottwald, C-H. Lambert, D.~Steil, V.~Uhl{\'\i}{\v r}, L.~Pang,
  M.~Hehn, S.~Alebrand, M.~Cinchetti, G.~Malinowski, Y.~Fainman,
  M.~Aeschlimann, and E.~E. Fullerton.
\newblock Engineered materials for all-optical helicity-dependent magnetic
  switching.
\newblock {\em Nat Mater}, 13(3):286--292, 03 2014.
\bibitem{ALyberatos}
A.~Lyberatos.
\newblock private communication.

\end{thebibliography}

\end{document}